\documentstyle[12pt]{article}
\newcommand{\be}{\begin{equation}}
\newcommand{\ee}{\end{equation}}
\newcommand{\s}{\section}
\newcommand{\ci}{\cite}
\newcommand{\r}{\ref}
\newcommand{\lb}{\label}

\begin{document}
\begin{center}

{\Large{\bf Hadron correlators and the structure of
the quark propagator  $^*$}}\\

\vspace{0.4cm}

{\large A. Ferrando $^{1}$}\\
{Departament de F\'{\i}sica Te\`{o}rica, Universitat de Val\`encia}\\
{and I.F.I.C., Centre Mixt Universitat de Val\`{e}ncia -- C.S.I.C.}\\
{E-46100 Burjassot (Val\`{e}ncia), Spain.}\\
\vspace{0.1cm}
{\large V. Vento $^{2}$}\\
{Dipartimento di Fisica, Universit\`a degli Studi di Trento}\\
{and I.N.F.N., gruppo collegato di Trento.}\\
{I-38050 Povo (Trento) Italy}\\
\vspace{0.4cm}
{\bf Abstract}
\begin{quotation}
{\small The structure of the quark propagator of $QCD$ in a confining
background is not known. We make an Ansatz for it, as hinted by a
particular mechanism for confinement, and analyze its implications
in the meson and baryon correlators. We connect the various
terms in the K\"allen-Lehmann representation of the quark
propagator with appropriate combinations of hadron correlators,
which may ultimately be calculated in lattice $QCD$. Furthermore,
using the positivity of the path integral measure for vector like
theories, we reanalyze some mass inequalities in our formalism.
A curiosity of the analysis is that, the exotic components of the
propagator (axial and tensor), produce terms in the hadron correlators
which, if not vanishing in the gauge field integration,
lead to violations of fundamental symmetries.
The non observation of these violations implies restrictions in
the space-time structure of the contributing gauge field configurations.
In this way, lattice $QCD$ can help us analyze the microscopic structure of
the mechanisms for confinement.}
\end{quotation}
\end{center}
\vspace{0.2cm}
$^*$Supported in part by CICYT (AEN91-0234) and DGICYT
grant (PB91-0119-C02-01). To be published in Z. Phys. C.\\
$^{1}$ ferrando@evalvx.ific.uv.es \\
$^{2}$ vento@evalvx.ific.uv.es.
On leave of absence
from Universitat de Val\`encia and I.F.I.C.
\newpage

\s{Introduction}
The structure of the quark propagator of $QCD$ in a confining
background has been a
matter of debate over the years \ci{vacuum}. The various
mechanisms for confinement hint different types of
vacua and therefore different quark propagators
\ci{vacuum,monopoles}. In particular the electric vortex
mechanism (see Appendix A) provides us with a quark propagator
with the following K\"allen-Lehmann representation,
\be
S(x,y) = s(x,y) + v_{\mu}(x,y)\gamma_{\mu} +
a_{\mu}(x,y) \gamma_{\mu} \gamma_5 + t_{\mu\nu}(x,y) \sigma_{\mu\nu}
\lb{propagator}
\ee
We shall take this as an Ansatz for the structure of the quark
propagator in a background to be used in the formalism of the so
called $QCD$ inequality approach.

It was realized long after $QCD$ was formulated that one could
derive some exact inequalities between hadron masses
\ci{inequalities} and other observables \ci{inequalities1}.
The key element in deriving them is that the
Euclidean fermion determinant in vector like gauge theories (such
as $QCD$) is positive definite and so the measure
\be
d\mu = Z^{-1} D A^a_{\mu} (x) det (i{\not\!\! D} + M)
\exp{(-\frac{1}{2g^2} \int d^4x TrF_{\mu \nu}^2)}
\ee
for the $A^a_{\mu}$ integration obtained after integrating out
the fermions is positive definite for $\Theta = 0$. Note that
${\not\!\! D} = \gamma^{\mu} D_{\mu}$, $D_{\mu}$ being the covariant
derivative. Inequalities
that hold pointwise continue to hold
after integrating with respect to a positive measure. Thus any
inequality among matrix elements that holds after performing the
Fermi integral in a fixed background gauge field holds in the
exact theory. The continuous formulation requires from an
appropriate regularization scheme \ci{asorey}.
The great advantage
of this procedure is that one sums over positive
contributions weighted by a positive measure and
therefore possible cancellations between different
gauge configurations are not worrysome.

The aim of this paper is to analyze various consequences of
Eq.(\ref{propagator}) within the inequality approach for $QCD$
\footnote{Note that our analysis may be applied to any
representation of the quark propagator. We have just chosen the
above equation, because it is hinted by a specific mechanism of
confinement and because it is sufficiently rich to allow the
most general analysis.}. Our
interest is twofold. On the one hand we shall discuss properties
of $QCD$, i.e., chiral symmetry realization, mass relations,...,
as if the above Ansatz were the outcome of the true calculation.
On the other we shall relate
the terms in the Ansatz to hadron correlators, which can
ultimately be calculated in lattice $QCD$. Finally we shall
discuss observable consequences of the exotic terms in the
K\"allen-Lehmann representation of the quark propagator, which
imply, to avoid violation of fundamental symmetries, a strong
restriction of the space-time structure of the contributing gauge
field configurations.

\s{The structure of the mesonic correlators}

Mass inequalities have been obtained among the mesons and
comparing baryons with mesons \ci{inequalities}. The important
property in these calculations has been
\be
S^+(x,y) =\gamma_5 S(y,x) \gamma_5
\label{gamma5}
\ee
where $S(x,y)$ is the quark propagator in a background. Our aim
is to discuss mass relations also among baryons. In this case
due to their current quark constituency the previous property
is of no use. Some fine details of the structure of the quark
propagator and of the baryon currents will be necessary to be
able to address the issue. We proceed thus with the same
technique in both cases, only that in the meson case we will use
of this simplification.

The meson correlators in terms of meson fields
are given by \footnote{Since global numerical factors
are of no relevance for our discussion, we later on
normalize the measure $d\mu$ so that the
coefficient of the scalar term is unity for the
trace correlator.}
\begin{eqnarray}
<\sigma(x) \sigma(y)> & = & -\int d\mu
Tr(\gamma_5 S^+(x,y) \gamma_5S(x,y))
\lb{sigma}\\
<\pi(x) \pi(y)> &= & \;\; \int d\mu
Tr(S^+(x,y)S(x,y))
\lb{pi}\\
<\rho_{\mu}(x) \rho_{\nu}(y)> &=& \;\; \int d\mu
Tr(\gamma_5S^+(x,y)\gamma_5\gamma_{\mu}S(x,y)\gamma_{\nu})
\lb{rho}\\
<\alpha_{\mu}(x) \alpha_{\nu}(y)>& =& - \int d\mu
Tr(S^+(x,y)\gamma_{\mu}S(x,y)\gamma_{\nu})
\lb{axial}\\
<\tau_{\mu \nu}(x) \tau_{\lambda \varphi}(y)>& = & - \int d\mu
Tr(\gamma_5S^+(x,y)\gamma_5
\sigma_{\mu \nu}S(x,y)\sigma_{\lambda \varphi})
\lb{tensor}
\end{eqnarray}
According to our previous discussion one should investigate
the properties of these correlators for the quark propagator
in the presence of a background field.
The substitution of Eqs.(\r{propagator}) and (\r{gamma5})
into Eqs.(\ref{sigma})
through (\r{tensor}) provides us with the structure of the
mesonic correlators. With the help of Mathematica and HIP
\ci{hip} the calculation is straightforward (see Appendix B).
We discuss here some of the properties of the arising
structures.

It was noticed some time ago \ci{nussinov} that the difference
between the sigma and pion correlators
\be
<\pi \pi> - <\sigma \sigma> = \int d\mu (|s|^2 + 2|t|^2)
\ee
could be non vanishing if anomalous structures were present
in the quark propagator. In such a case chiral symmetry would
be broken by the mechanism leading to these structures. In our
case the electric vortex contributes both to $s$ and
$t_{\mu \nu}$ leading to the spontaneous breaking of chiral
symmetry. However this contributions are proportional to the
fermion mass, thus our mechanism could never explain the
spontaneous breaking of chiral symmetry in a massless theory.
However this is not an inconvinience, since as can be seen
in ref.\ci{vafa}, the mass plays also a crucial role in more
fundamental approaches. In the present model the restoration
of chiral symmetry and deconfinement would occur at the same
scale. However this statement has to be taken with precaution
due to our simplified scenario. It could happen that other
mechanisms, like for example instanton effects, could modify
the conclusions \ci{shuryak}.

In our formalism correlator inequalities leading to mass relations
can be constructed in an explicit fashion.
One can easily see that the pion has the lowest mass since the
right hand sides of the following equations are positive definite
\begin{eqnarray}
<\pi \pi> - <\sigma \sigma>& = & 2 \int d\mu (|s|^2 + 2|t|^2)\\
<\pi \pi> - \frac{1}{4}<\rho \rho>& = &\frac{1}{2} \int d\mu
(|v|^2 + 3 |a|^2)\\
<\pi \pi> - \frac{1}{4}<\alpha \alpha>& =& 2 \int d\mu
(|s|^2 + \frac{1}{2} |v|^2 + \frac{3}{2} |a|^2)\\
<\pi \pi> - \frac{1}{12}<\tau \tau>& =& \;\; \int d\mu
(|v|^2 + |a|^2 + 3|t|^2)
\end{eqnarray}
Moreover one can isolate from the different correlators the
various contributions. In particular the scalar term simply
states that the axial meson has a bigger mass than the vector
meson
\be
<\rho \rho> -<\alpha \alpha> = 8 \int d\mu |s|^2
\lb{vector}
\ee
The rest are less instructive
\be
<\pi \pi> +\frac{1}{8}<\rho \rho> + \frac{3}{8} <\alpha \alpha>
= \;\; \int d\mu |v|^2
\ee
\be
<\pi \pi> - \frac{3}{8}<\rho \rho> - \frac{1}{8}<\alpha \alpha>
= 2\int d\mu|a|^2
\ee
\be
<\rho \rho> - \;\;<\alpha \alpha> - \frac{2}{3}<\tau \tau> =
8\int d\mu |t|^2
\ee
The last equation, together with Eq.(\r{vector}), tell us
that the vector meson mass is smaller than the tensor meson
mass. The remaining confirm the fact that the pion has the
lowest mass among the mesons.

A corolary of our relations is that by studying the mesonic
correlators one may be able to disentangle the existence or
non-existence of anomalous terms. Therefore one should try
to understand the structure of the full quark propagator
{\em experimentally} (lattice $QCD$).

Once this structures have been unveiled the analysis of the
various terms in the correlators becomes very rich.
In first place terms with $\varepsilon_{\mu \nu \rho \sigma}$
appear which have to vanish after gluon integration if the theory
is to be Poincar\'e invariant. This implies already a strong
restriction on the space-time structure of the contributing gauge
fields. However in the tensor meson correlator
contributions of the form
\be
\int d\mu \varepsilon_{\mu \nu \rho \sigma} ( a^* \cdot v
+a \cdot v^*)
\ee
\be
\int d\mu \varepsilon_{\mu \nu \rho \alpha} (a^*_{\alpha}  v_{\rho}
\pm a_{\alpha} v^*_{\rho})
\ee
\be
\int d\mu \varepsilon_{\mu \nu \rho \alpha} (a_{\rho} v^*_{\alpha}
\pm a^*_{\rho} v_{\alpha})
\ee
arise, which are perfectly compatible with Poincar\'e invariance,
and if non vanishing, signal the violation of $CP$ and $P$
invariance. If the latter are not observed, as it seems, or very
small, then this will also imply further restrictions on the
space-time structure of the allowed gauge field configurations.

\s{ The structure of the baryonic correlators}
The first step in our development is the construction of the
baryon currents from the constituents. We shall restrict
ourselves to composite operators with baryon quantum
numbers and with the least possible dimension. This leads
to currents proportional to the fields without derivatives
\ci{ioffe}.

The quark fields are Dirac spinors and therefore belong to
the ${\cal D}^+_{\frac{1}{2},0}$ representation of the Lorentz
group, where
\be
{\cal D}^+_{\frac{1}{2},0} = {\cal D}_{\frac{1}{2},0}
\oplus {\cal D}_{0,\frac{1}{2}}
\ee
The baryon currents are obtained by reducing the product
of three Dirac fields
\be
\Psi^{fa}_{\alpha}\otimes\Psi^{gb}_{\beta}
\otimes\Psi^{hc}_{\gamma}
\ee
where $\alpha,\beta,\gamma$ denote the spinor, $a,b,c$ the color
and $f,g,h$ the flavor indices. The reduction is not a trivial
exercise \ci{t-rafecas}. Our result, reproducing that of Dosch
et al. \ci{dosch}, is

Proton:
$$
A u^T(x) {\cal C} \gamma_5 d(x) \gamma_{\mu} u_{\lambda}(x) +
B u^T(x) {\cal C} d(x) \gamma_5\gamma_{\mu} u_{\lambda} (x)
$$
\be
+ C u^T(x) {\cal C} \gamma_5 \gamma_{\rho} d(x)
(\delta^{\mu \rho} - \frac{1}{4} \gamma^{\mu}\gamma^{\rho})
u_{\lambda}(x)
\ee

Delta$^{++}$:
\be
Du^T{\cal C}\gamma^{\mu}u\gamma^{\nu}u_{\lambda} +
Eu^T{\cal C}\sigma^{\mu \nu} u u_{\lambda}
\ee
where A,B,... are independent constants. The argument of
Espriu et al. \ci{espriu} is that in order to preserve
the same order in momentum, C=E=0, since the projection
operator to the ${\cal D}^+_{\frac{3}{2},0}$ representation
depends on momentum, a statement which is certainly true
in the free case. Accepting this argument one recovers
Ioffe's result which can be rewritten by appropriate
Fierzing as \ci{espriu}

Proton:
\be
u^T{\cal C}\gamma_5du_{\lambda} + \xi u^T{\cal C}d\gamma_5u_{\lambda}
\ee

Delta$^{++}$:
\be
u^T{\cal C}\gamma^{\mu}uu_{\lambda}
\ee
with the caveat that although $\xi$ is in principle arbitrary,
in the case of chiral symmetry it has the value $\xi = -1$.

The calculation of the correlators gives for the proton
\be
<P_{\mu}(x){\bar P}_{\nu}(y)> = <>_{11} + \xi(<>_{12} + <>_{21})
+ \xi^2 <>_{22}
\ee
where
\begin{eqnarray}
<>_{11}& = &\int d\mu \{(\gamma_5 S(x,y) {\cal C}S^T(x,y){\cal C}
S(x,y)\gamma_5)_{\mu \nu} - Tr(S(x,y){\cal C}S^T(x,y){\cal C})
(\gamma_5S(x,y)\gamma_5)_{\mu \nu}\} \nonumber \\
<>_{12}& =& \int d\mu \{(S(x,y) {\cal C}S^T(x,y){\cal C} \gamma_5
S(x,y)\gamma_5)_{\mu \nu} - Tr(S(x,y){\cal C}S^T(x,y){\cal C}
\gamma_5) (S(x,y)\gamma_5)_{\mu \nu}\} \nonumber\\
<>_{21}& =& \int d\mu \{(\gamma_5 S(x,y) \gamma_5
{\cal C}S^T(x,y){\cal C}S(x,y))_{\mu \nu} -
Tr(S(x,y)\gamma_5{\cal C}S^T(x,y){\cal C})
(\gamma_5S(x,y))_{\mu \nu}\} \nonumber\\
<>_{22}& = &\;\; \int d\mu \{(S(x,y) \gamma_5 {\cal C}S^T(x,y){\cal C}
\gamma_5 S(x,y))_{\mu \nu} - Tr(S(x,y) \gamma_5 {\cal C} S^T(x,y)
{\cal C}\gamma_5) S(x,y)_{\mu \nu}\} \nonumber \\
& &
\end{eqnarray}
and for the Delta$^{++}$
$$
<\Delta^{\alpha}_{\mu}(x) {\bar \Delta}^{\beta}_{\nu}(y)> =
\int d\mu \{2(S(x,y)\gamma^{\beta}{\cal C}S^T(x,y){\cal C}
\gamma^{\alpha} S(x,y))_{\mu \nu} \;\;\;\;\;\;\;\;\;\;\;\;
$$
\be
\;\;\;\;\;\;\;\;- Tr(S(x,y)\gamma^{\beta}
{\cal C} S^T(x,y){\cal C}\gamma^{\alpha})S(x,y)_{\mu \nu}
\ee

We next take the equation for the propagator
Eq.(\r{propagator}) into the equations of the baryonic
correlators. The calculation can be easily
performed with Mathematica and HIP \ci{hip} but the result is too
messy to be shown here (in Appendix C we show some of the terms
for the proton). We proceed to discuss certain
features which are relevant.

Let us discuss the mass relations. The diagonal correlators
are given by
\be
\frac{1}{4} <P{\bar P}> = \int d\mu \{s(s^2 - \frac{1}{2} v^2
+ \frac{1}{2} a^2 - 2 t^2) + \varepsilon_{\mu \nu \lambda \varphi}
a_{\mu}v_{\nu} t_{\lambda \varphi}\}
\ee
and
\be
\frac{1}{16}<\Delta {\bar \Delta}> = \int d\mu
\{s(s^2 + v^2 -  a^2 + \frac{2}{3} t^2) +
\frac{2}{3}\varepsilon_{\mu \nu \lambda \varphi}
a_{\mu}v_{\nu} t_{\lambda \varphi}\}
\ee

It is immediate to realize that the terms on the right hand
side are individually not of definite sign and therefore the
possibility of obtaining adequate relations leading to mass
inequalities diminishes considerably. An appropriate choice
which leads after repeated use of H\"older's and Schwarz'
inequalities to a meaningful bound is
\be
|\frac{1}{6}<P{\bar P}> - \frac{1}{16}<\Delta{\bar \Delta}>|
\leq |<\pi \pi>|^{\frac{3}{2}}
\lb{bound}
\ee
and therefore the following bound arises
\be
m_{Baryon} \geq \frac{3}{2} m_{\pi}
\ee
We cannot say however which of the two baryons is the heaviest,
since all meaningful bounds imply modulus of differences of
the correlators like in Eq.(\r{bound}).

If one looks at the full correlators,
anomalous terms (pseudoscalar, axial and tensor) appear both in the
the proton and the delta correlators. The tensor terms should vanish
after integration if Poincar\'e invariance is to hold. The
pseudoscalar and axial, if not vanishing after integration, signal
a violation of $CP$ and $P$ in the strong interaction
\ci{rafecas}.

\s{Conclusion}

Within a plausible scenario for confinement
\ci{mandelstam,giacomo}
we have discussed the most general possible structure of the
quark propagator.
Using a current description for the hadrons we have calculated
their correlators, limiting ourselves to those corresponding
to low lying hadrons \footnote{Recently \ci{manohar} similar
techniques have been applied to heavy quark baryons.}. Our
calculation has remained qualitative because we have not
been able, despite many efforts, to find a solvable model
\'a la Schwinger. However we have laid down the formalism
for a quantitative analysis via lattice calculations. Furthermore
the formalism is independent of the particular structure chosen
for the quark propagator.

We have recalculated some of the correlator inequalities
leading to mass relations and chiral symmetry realization
in an effort to show the contribution to these observables
arising from the so called anomalous terms. In this way
we have related these terms to hadron-hadron correlators
which are in principle calculable.

We have shown explicitly that the appearence of anomalous
terms is not a peculiarity of the formalism, but a quite
general phenomena of Schwinger's equations.
The existence of pseudoscalar, axial and tensor
terms  has important implications from the physical
point of view. They may be instrumental in describing the
realization of chiral symmetry \ci{nussinov} and moreover
might lead to observable consequences associated with the
violation of discrete symmetries \ci{rafecas}.

To eliminate these anomalous components from the fermion propagators
before gluon integration one needs structureless color fields.
The elimination through the integration requires high
gluon entropy. The observed feature of none or very small
violation of discrete symmetries by the strong interaction
implies necessarily a strong dynamical restriction on the
possible confinement mechanisms. It would be very advisable
to have, no matter how naive, a solvable model that would
instruct us on how these facts restrict the
structure of the color fields. In this way we could forsee
questions to be asked to the more exact, but less intuitive
lattice calculations. In the meantime we have to resort to
qualitative features and hopefully {\em experimental}
observations.

\s*{Acknowledgement}
It is a pleasure to thank Arcadi Santamaria. Many of the
calculations would not have been possible without his help.
Correspondence with Stuart Samuel was most useful.
Domenec Espriu  resucitated some of his old notes
providing us with valuable insight into his calculation.
Useful discussions with Magda Rafecas are acknowledged.
We would like to thank Rafael Guardiola for his assistance, which
made the use of Kate and Spark (Unix operating systems) possible.

\newpage
\appendix
\begin{center}
{\bf\LARGE Appendices}
\end{center}

\s{The structure of the quark propagator in an electric vortex}

The confinement phase of $QCD$ can be understood as a
coherent plasma of monopoles \ci{mandelstam,thoofta}.
This phase allows for electric vortices
and therefore color charges are confined.
In spite the appeal of this proposal no one has been able
to use this characterization for realistic quantitative
calculations in continuum $QCD$. However a topological
description of confinement has arisen \ci{stuart} which
has been implemented \ci{thooftb,luescher} and searched
for in in lattice calculations
\ci{giacomo}.

Let us accept this appealing scenario and assume that a
hadron is represented simply by two {\em opposite}, i.e.
$N$ and ${\bar N}$, $N$ being the number of colors,
color charges connected by a color
electric vortex. This scenario is no more than the dual
of the confinement mechanism in $QED$ once monopoles
are included in the theory. In this case the monopole
and antimonopole are confined by a magnetic vortex
\ci{mandelstam,nielsen}. Thus a hadron consists of a
string like configuration between two opposite charges,
which in the case of (non exotic) mesons are a quark and
an antiquark, and in the case of (non exotic) baryons
a quark and a diquark. This {\em hadronic} configurations
exist on top of a highly non perturbative vacuum, the
monopole plasma, that can be understood as a very disordered
(large entropy) system of color magnetic flux tubes \ci{olesen}.
This latter description of the vacuum, as a disordered system,
motivates our second assumption, namely that local magnetic
effects will disappear in the averaging process,i.e., on a
global scale where observable effects take place. Thus
from the observational point of view the role of the
complicated non perturbative vacuum is just to allow for the
confinement scheme sketched above. We do not take into
account other non perturbative mechanisms arising from
a more complete description of the vacuum, e.g., instantons,
which might contribute also to some of the effects we shall
discuss.

In order to obtain the structure of the quark propagator in the
vecinity of a color electric vortex
we repeat the construction of the Nielsen-Olesen vortex
\ci{nielsen} for the dual fields obtaining an electric vortex
of the form
\be
\vec{E} = \hat{z} f(x_1,x_2)
\ee
The color indices saturate with the appropriate choice of the
color structure of the non abelian vortices and drop out of the
calculation.

In Schwinger's \ci{schwinger} proper time method the equations
of motion become
\be
\frac{d\pi_1}{ds} = \;\; g \frac{\partial f(x_1,x_2)}
{\partial x_1} \;\sigma_{34}  \;\;\;\;\;
; \;\;\;\; \frac{dx_1}{ds} = 2\pi_1
\lb{schwinger1}
\ee
\be
\frac{d\pi_2}{ds} = \;\; g\frac{\partial f(x_1,x_2)}
{\partial x_2}\;\sigma_{34}   \;\;\;\;\;
; \;\;\;\; \frac{dx_2}{ds} =
2\pi_2
\lb{schwinger2}
\ee
\be
\frac{d\pi_3}{ds} = -2g  f(x_1,x_2) \;\pi_4
\;\;\;\;\;\;\; ; \;\;\;\; \frac{dx_3}{ds} = 2\pi_3
\ee
\be
\frac{d\pi_4}{ds} = \;\; 2g  f(x_1,x_2) \; \pi_3
\;\;\;\;\;\;\;\; ; \;\;\;\; \frac{dx_4}{ds} = 2\pi_4
\lb{schwinger4}
\ee
It is possible to obtain first integrals of these
equations, i.e.,
\begin{eqnarray}
\frac{d}{ds} (\pi^2_1 + \pi^2_2) & = & g
\frac{df(x_1,x_2)}{ds} \;\sigma_{34} \\
\frac{d}{ds} (\pi^2_3 + \pi^2_4) & = & 0
\end{eqnarray}
that lead to a hamiltonian for the evolving quasi particle
\be
{\cal H}(s) = \pi^2 (0) - g\sigma_{34} f(x_1(0),x_2(0))
\lb{hamiltonian}
\ee
The first of Schwinger's equations reads
\be
i\partial_s(x(s)'|x(0)'') = (x(s)'|{\cal H}|x(0)'')
\lb{schrodinger}
\ee
Using Eq.(\r{hamiltonian}) we can rewrite Eq.(\r{schrodinger})
as
\be
i\partial (\tilde{x}(s)'|\tilde{x}(0)'') =
(\tilde{x}(s)'|\pi^2 (0)|\tilde{x}(0)'')
\lb{spinless}
\ee
just by rotating in spin space as
\be
|x(s)) =
e^{-ig\sigma_{34}f(x_1(0)'', x_2(0)'') s} |\tilde{x}(s))
\ee

Equation (\r{spinless}) corresponds to that of a spinless field
and can be integrated formally together with the remaining
equations of Schwinger \ci{schwinger} leading to \ci{vafa}
$$
D^A(x',x'',m) = \int^{\infty}_0 ds \int^{x(s)=x''}_{x(0)=x'}
[dx^{\mu}] \exp{(-i\int^s_0ds(\frac{dx^{\mu}}{ds})^2)}\;\;\;
$$
\be
\;\;\;\;\;\;\;\;\;\;\;\;\;\;\;
\exp{(-im^2s)} {\cal P} (\exp{(i\int^{x(s)=x''}_{x(0)=x'}A_{\mu}
dx^{\mu})})
\lb{boson}
\ee
In particular the last factor is a consequence of the
additional proper time equations \ci{schwinger}.
One obtains the fermionic propagator from
Eq.(\r{boson}) in a straightforward fashion
$$
S^A(x',x'',m) = \int^{\infty}_0 ds \int^{x(s)=x''}_{x(0)=x'}
[dx^{\mu}] (\gamma_{\mu} \frac{dx^{\mu}}{ds} + m )
\exp{(-i\int^s_0ds(\frac{dx^{\mu}}{ds})^2)}\;\;\;\;
$$
\be
\;\;\;\;\;\;\;\;\;\;\;\;
\exp{(-im^2s)} \exp{(ig\sigma_{34}f(x'',x')s)}
{\cal P} (\exp{(i\int^{x(s)=x''}_{x(0)=x'}A_{\mu}
dx^{\mu})})
\lb{fermion}
\ee
Since $\sigma_{34}^2 = 1$  the spin phase becomes
\be
e^{ig\sigma_{34}f(x'',x') s} = \cos{(gf(x'',x')s)} +
i\sigma_{34}\sin{(gf(x'',x')s)}
\ee
Furthermore
\be
\gamma_{\mu}\sigma_{34} = i(\delta_{\mu 3} \gamma_4
- \delta_{\mu 4}\gamma_3) + \varepsilon _{34\mu \nu}
\gamma_5 \gamma_{\nu}
\ee
Thus the fermion propagator will contain besides the
conventional scalar and vector terms, axial and tensor
terms.

\s{Some Meson correlators}
In this appendix we show some of the non trivial meson
correlators, in particular the vector and axial mesons,
since the scalar and pseudoscalar meson correlators are given
in the text and since the tensor meson correlator is too messy,
we shall only show some of its anomalous terms.

For the vector meson we have
\begin{eqnarray}
<\rho_{\mu} \rho_{\nu}> & = &\int d\mu
\{ \delta_{\mu \nu} (|s|^2 + |v|^2 - |a|^2 + 2 |t|^2)
- (v_{\mu}v^*_{\nu} + v^*_{\mu} v_{\nu}) \nonumber \\
& & + (a_{\mu}a^*_{\nu} + a^*_{\mu} a_{\nu})
+ \varepsilon_{\mu \nu \lambda \varphi}
(a^*_{\lambda} v_{\varphi} + a_{\lambda} v^*_{\varphi}) \nonumber \\
& & 2i(s^*t_{\mu \nu} - s t^*_{\mu \nu})
+ 4 (t_{\mu \lambda} t^*_{\lambda \nu} +
t_{\nu \lambda} t^*_{\lambda \mu})\}
\end{eqnarray}
and for the axial meson
\begin{eqnarray}
<\alpha_{\mu} \alpha_{\nu}> & = &\int d\mu
\{ \delta_{\mu \nu} (- |s|^2 + |v|^2 - |a|^2 - 2 |t|^2)
- (v_{\mu}v^*_{\nu} + v^*_{\mu} v_{\nu}) \nonumber \\
& & + (a_{\mu}a^*_{\nu} + a^*_{\mu} a_{\nu})
+ \varepsilon_{\mu \nu \lambda \varphi}
(a^*_{\lambda} v_{\varphi} + a_{\lambda} v^*_{\varphi})
\nonumber \\
& &- 2i(s^*t_{\mu \nu} - st^*_{\mu \nu})
- 4 (t_{\mu \lambda} t^*_{\lambda \nu} +
t_{\nu \lambda} t^*_{\lambda \mu})\}
\end{eqnarray}
We notice that in these cases the $CP$ and $P$ violating terms
must vanish after integration if Poincar\'e invariance is
to hold.

The tensor meson contains the following terms which are not
forced to vanish after integration by Poincar\'e invariance
\begin{eqnarray}
<\tau_{\mu \nu} \tau_{\lambda \varphi}>& = & \int d\mu \{
\varepsilon_{\mu \nu \lambda \varphi}
(a \cdot v^* + a^* \cdot v)
+ \varepsilon_{\mu \nu \lambda \eta}
(a^*_{\varphi} v_{\eta} -a_{\varphi} v^*_{\eta}) \nonumber \\
& & + \varepsilon_{\mu \lambda \varphi \eta}
(a^*_{\nu} v_{\eta} + a_{\nu} v^*_{\eta})
- \varepsilon_{ \mu \nu  \varphi \eta}
(a^*_{\lambda} v_{\eta} -a_{\lambda} v^*_{\eta}) \nonumber \\
& & - \varepsilon_{\nu \lambda \varphi \eta}
(a^*_{\mu} v_{\eta} + a_{\mu} v^*_{\eta})
- \varepsilon_{ \nu \lambda \varphi \eta}
(a^*_{\eta} v_{\mu} + a_{\eta} v^*_{\mu}) \nonumber \\
& & + \varepsilon_{ \mu \nu \varphi \eta}
(a^*_{\eta} v_{\lambda} - a_{\eta} v^*_{\lambda})
+ \varepsilon_{ \mu \lambda \varphi \eta}
(a^*_{\eta} v_{\nu} + a_{\eta} v^*_{\nu}) \nonumber \\
& & - \varepsilon_{ \mu \nu \lambda  \eta}
(a^*_{\eta} v_{\varphi} - a_{\eta} v^*_{\varphi})
+ \ldots \}
\end{eqnarray}

\s{Some anomalous terms of the proton correlator}
In the chiral limit ($\xi = -1$) the following anomalous
terms arise in the proton correlator
\begin{eqnarray}
<P {\bar P}> & = & \int d\mu \{
i a_{\alpha} v_{\beta} t_{ \beta \alpha} \gamma_5
+ [(\frac{1}{2}s^2 + v^2 + a^2 + 2t^2)a_{\alpha} \nonumber \\
& & -2(a \cdot v) v_{\alpha} + 4 t_{\alpha \lambda}
t_{\lambda \varphi} a_{\varphi}
-i\varepsilon_{\alpha \beta \lambda \varphi}t_{\beta \lambda}
v_{\varphi}]\gamma_5 \gamma_{\alpha} \nonumber \\
& & + [\frac{i}{2} s \varepsilon_{\alpha \beta \lambda \varphi}
a_{\lambda} v_{\varphi}
+ \frac{1}{2} \varepsilon_{\beta \lambda \varphi \sigma}
(a_{\alpha} t_{\lambda \varphi} v_{\sigma} +
a_{\lambda} t{\alpha \sigma} v_{\varphi} +
a_{\lambda}t_{\sigma \varphi} v_{\alpha}) \nonumber \\
& & + \frac{1}{2}\varepsilon_{\alpha \lambda \varphi \sigma}
(a_{\lambda} t_{\sigma \beta} v_{\varphi} +
a_{\beta} t{\varphi \lambda} v_{\sigma} +
a_{\lambda}t_{\varphi \sigma} v_{\beta})
+ (s^2 + \frac{1}{2}v^2 - \frac{1}{2} a^2 + 2 t^2)
t_{\alpha \beta} \nonumber \\
& & + a_{\alpha} a_{\lambda} t_{\lambda \beta}
+ a_{\beta}a_{\lambda}t_{\alpha \lambda}
+ v_{\beta} v_{\lambda} t_{\beta \lambda}
+ v_{\alpha} v_{\lambda} t_{\lambda \alpha}]\sigma_{\alpha \beta}
+ \ldots \}
\end{eqnarray}

\end{document}